\documentclass[a4paper,11pt]{article}
\usepackage[english]{babel}
\usepackage[utf8]{inputenc} %utf8x
\usepackage{amsmath,amssymb}
\usepackage{graphicx}
\usepackage[colorlinks=true, allcolors=black]{hyperref}
\usepackage{geometry}
\usepackage{natbib}
\usepackage{graphics}
\usepackage{setspace}

\begin{document}

%Redistributive Policies and Transformative AI
%Redistributive Policies for the Times of Technological Singularity
\title{Redistributive Policies for the Times of Transformative AI\thanks{We would like to thank the participants of the 14th Summer Workshop on Macroeconomics and Finance (Warsaw, Poland) for their helpful comments and suggestions. We acknowledge the research funding from the National Science Center, Poland, through the grant ``Will Artificial General Intelligence Bring Extinction or Cornucopia? Modeling the Economy at Technological Singularity'' (OPUS 26 No. 2023/51/B/HS4/00096).}}

\author{Jakub Growiec\thanks{SGH Warsaw School of Economics, Poland and CEPR Research and Policy Network on AI.}  \and Klaus Prettner\thanks{WU Vienna, Austria.} \and Maciej Szkr{\'o}bka\thanks{SGH Warsaw School of Economics, Poland}}

\maketitle

%\begin{center}
%Preliminary draft
%\end{center}

\vspace{20mm}

\begin{abstract} 
\noindent After the arrival of transformative artificial intelligence (TAI), broad-based automation is expected to decrease the labor share and increase income and wealth inequality. Although economic growth is likely to accelerate, most of its gains may accrue to a narrow group of individuals and firms. Hence, if unmitigated by redistributive policy, income and wealth inequality may rise to levels unseen in the industrial economy. Using a unifying theoretical framework, we survey the redistributive policies proposed for the era of TAI, such as universal basic income (UBI), universal basic capital (UBC), state-issued compute and robot permits, and taxes on capital, compute, robots, tokens, land, energy, consumption, and wealth. We argue that policies that are able to broadly distribute rents from capital including compute and robots---such as UBC or UBI financed through capital taxes---are most likely to achieve lasting reductions in inequality in a world with human-aligned transformative AI.

\bigskip

\noindent \textbf{JEL codes:} H20, J30, O11, O33. 

\noindent \textbf{Keywords:} Transformative Artificial Intelligence (TAI), Redistributive Policy, Universal Basic Income, Technological Singularity. 
\end{abstract}

\newpage

\section{Introduction}

Artificial intelligence is on the rise; this broad claim refers both to AI inputs and measurable AI capabilities. Cumulative computing power allocated to the training and inference of frontier AI models is doubling every six months \citep{epoch2026aimodels}. Due to rapid growth in AI capabilities across all considered narrow domains (such as reading comprehension, image recognition, predictive reasoning, etc.), all existing benchmarks are quickly saturated \citep{owid-artificial-intelligence}. And today's AI models are no longer passive chatbots either, as indicated by their rising agentic capabilities: the length of tasks AI can reliably perform is doubling every 7 months \citep{METR}. Evidence has already been provided that modern-day multimodal language and reasoning models can be classified as artificial general intelligence (AGI), i.e., a type of AI that matches or surpasses human capabilities across a wide range of cognitive tasks \citep{Norvig2023,Aguera2025,NatureAGI}.

Despite this impressive growth, the economic impacts of AI beyond narrow applications (such as coding or telemarketing) have hitherto been disappointing \citep[see e.g.,][]{Acemoglu2025Simple,BonfiglioliEtal2025}. This is likely because of adoption lags, the ``J-curve'' along which productivity often dips after the introduction of a new technology before it eventually increases \citep{BrynjolfssonEtal2017}, and the need for complementary investments and managerial decisions necessary for reorganizing corporate workflows \citep{AgrawalEtal2022}. Nonetheless, given that generative AI is a general-purpose technology able to automate human cognitive labor at scale, we might soon encounter a massive socio-economic transformation as suggested, for example, by \cite{KorinekStiglitz2019}, \cite{AI2027}, \cite{Amodei2026} and \cite{KorinekEtal2026}. 

Transformative AI (henceforth TAI) can be defined as a suite of AI algorithms allowing unaided machines to perform every economically valuable task better than humans and thereby, if implemented, fully automate human labor. In the words of \cite{Karnofsky2016}, TAI would be an ``AI that precipitates a transition comparable to (or more significant than) the agricultural or industrial revolution.'' While such a transformation has not occurred yet, it constitutes a plausible scenario for the future which merits the preparation of a targeted policy response.

The broad intelligence and general applicability of prospective TAI will allow it to automate all essential tasks in the economy and thereby become substitutable, rather than complementary, to human cognitive work---a key property distinguishing TAI from any ``normal technology'' \citep{NormalTech}. From that point on, it is predicted that as AI capabilities increase further and the costs of operation go down, the labor income share will fall precipitously \citep{Growiec2020,Moll2022,GrowiecEtal2024,TrammellPatel2025,MinnitiEtal2025}. Whether this will be associated with rising technological unemployment depends on the properties of labor supply (for example, some individuals may be willing to work for any wage) and progress in complementary technologies such as robotics,\footnote{Shortage of TAI-operated physical actuators may temporarily benefit manual workers at the expense of cognitive ones. It is possible that it will take longer to replace human hands than brains.} but relative impoverishment of labor as compared with the owners of capital and AI appears highly plausible. In an economy undergoing a precipitous fall in the labor share, wages will gradually lose its role as the primary force of income distribution among the human population---highlighting that alternative distribution mechanisms ought to be implemented \citep[see also][chapter 7, for an overview]{PrettnerBloom2020_adapted}.

Simultaneously, given that human labor is naturally distributed---brains are not accumulable \emph{per capita}---whereas capital ownership tends to be strongly concentrated, and ownership of AI services even more so, this development will by default increase inequality. Absent any redistributive policy, the gains of accelerated economic growth will be likely captured by a narrow group of individuals \citep[][]{Berg2018}, such as the owners of AI labs and data centers, or even administered by the TAI itself. What can be done to mitigate those worrying trends and allow the majority of the human population, i.e., those who do not currently hold any significant amounts of capital, to benefit from AI-driven growth as well?

The contribution of this paper to the literature is a survey and categorization of redistributive policies that could assist the deep technological transformation of the global economy from the initial state in which the economic impacts of AI are modest, all the way to an economy where all essential production, R\&D, and decision making is carried out by TAI. We review the policy proposals put forward in the associated literature, with the goal of categorizing their key characteristics and expected consequences. %Using a unifying theoretical framework and d
Departing from a baseline \emph{laissez-faire} setup of no policy intervention, we demonstrate how each of the proposals could affect the factor income distribution and overall inequality among the human population over a sequence of increasingly transformative scenarios: % in an era of TAI. %and to contrast their implications in a 
(a) AI only as a tool in people's hands, (b) full automation of a fraction of essential tasks, (c) full automation of all essential tasks by TAI. We organize our results using a simple theoretical framework which defines the technology while deliberately underspecifying preferences and market design.

Our key observation is that in a world with human-aligned TAI, the key growth bottleneck, i.e., the slowest growing essential production factor, will be the physical capital complementary to TAI software---such as compute and robots. The income share of this factor will gradually rise towards unity. Hence, only policies targeting the ownership and rents from such forms of capital will be effective in reducing income inequality in a world with TAI. Such policies include universal basic capital (UBC) as well as universal basic income (UBI) indexed to capital income and financed from capital taxes.

The remainder of the paper is structured as follows. In Section \ref{sec:theory} we set up our theoretical framework organizing the results. In Section \ref{sec:policies} we survey the proposed redistribution policies. Section \ref{sec:concl} concludes.

%\textcolor{red}{[more here]}

\section{Theoretical Framework}\label{sec:theory}

We consider the development of AI a technological phenomenon with transformative economic consequences. To study these consequences up to the first order, we use a theoretical framework which carefully defines the technology while leaving preferences and market design deliberately underspecified. This allows us to focus on the most robust impact channels; more specific channels and secondary feedback mechanisms are left for further research.  

\subsection{The Supply Side of the Economy}

Consider an economy in which final output is produced using physical capital $K$, human cognitive work $H$, and $AI$. Physical capital encompasses all the machines which perform productive physical actions. Each of these machines requires to be operated either by a human operator or %can work autonomously, following some pre-programmed code or instructions fed 
by AI. Therefore, apart from traditional equipment and structures, physical capital includes also digital hardware (compute and robots).\footnote{This is a simplified version of the hardware--software framework developed by \cite{GrowiecEtal2024}. %We could also say that, to simplify, physical capital $K$ includes the physical actions performed by humans themselves, by domesticated animals, and even the productive plant growth on cultivated land.
} We assume that the actions performed by physical capital are complementary to the cognitive input provided by either humans or AI:
\begin{equation}\label{eq:F}
Y = F(K, G(H, AI)).
\end{equation}
With regard to the aggregation of human and AI instructions into the overall cognitive input $S=G(H, AI)$, we allow three possibilities. In the order of rising AI capabilities, we expect humans and AI to be either (i) broadly complementary to AI (humans using AI as a tool), (ii) complementary in some tasks but substitutable in others (when some tasks can be performed fully autonomously, but others cannot), or (iii) substitutable across all essential tasks (after the arrival of TAI). 

One could exemplify these assumptions, for example, using a two-level CES function with constant returns to scale:
\begin{equation}
Y = \left( \alpha K^\theta + (1-\alpha) S^\theta\right)^\frac{1}{\theta}, \qquad \theta<0,
\end{equation}
where $\theta<0$ implies that $K$ and $S$ are gross complements, and 
\begin{equation}
S= \left( \beta H^\omega + (1-\beta) AI^\omega\right)^\frac{1}{\omega}, \qquad \omega\in(-\infty,0)\cup(0,1].
\end{equation}
If $\omega<0$, we are in the pre-TAI world where AIs are tools in people's hands---or ``bicycles for the mind'' \citep{AgrawalEtal2025}. Conversely, if $\omega\in(0,1]$, we are in a world with TAI, where all essential tasks can be performed by the TAI alone, and the human contribution is inessential and replaceable \citep{KorinekSuh2024}. We also consider a transition period where in a fraction of essential tasks $\kappa\in(0,1)$, human input is necessary (so for these tasks, $\omega_\kappa<0$), whereas the remaining fraction $1-\kappa$ of essential tasks can be fully automated (for them, $\omega_{1-\kappa}\in(0,1]$), cf. \cite{AcemogluRestrepo2018,AcemogluRestrepo2019b}.

\subsection{Factor Ownership and Inequality}

To study the evolution of factor distribution and inequality, we need to first specify factor ownership. We assume that a fraction of physical capital $\zeta_K\in[0,1]$ is foreign-owned, so that the proceeds from capital rental are transferred abroad; analogously, $\zeta_{AI}\in[0,1]$ is the fraction of foreign-owned AI. In turn, the domestically owned production factors $X\in\{K, H, AI\}$ are distributed across the domestic population $i=1,...,N$ according to a certain distribution $\{x_i\}_{i=1}^N$ with mean $\bar x=X/N$ and variance $\sigma^2(x)$, such that $\sum_{i=1}^N x_i=X$.

We do not consider the symmetric possibility that the domestic economy obtains proceeds from capital invested abroad, or domestically owned AI applied abroad. This is because in what follows we are going to consider the policy proposals from three perspectives: (i) global policy (with $\zeta_K=\zeta_{AI}=0$), (ii) domestic policy in a country behind the technology frontier, (iii) policy in an economy partially controlled by autonomous TAI. Interestingly, from the domestic inequality perspective, the last two cases are entirely identical: for the domestic population it does not matter whether certain gains are appropriated by foreign nationals or autonomous AI.

We analyze inequality either from the factor endowment perspective, or from the disposable income perspective. 

\subsubsection{Distribution of Endowments}
Decomposing the production function \eqref{eq:F}, and using constant returns to scale, we obtain
\begin{equation}\label{eq:Y}
\ln Y = \pi_K \ln K + \pi_H \ln H + \pi_{AI} \ln AI,
\end{equation}
where $\pi_K, \pi_H$, and $ \pi_{AI}$ are the respective factor income shares which sum up to 1. Under perfect competition, these shares would be equal to the factors' respective marginal elasticities. For example, in the two-level CES case we would have:
\begin{equation}
\pi_K = \alpha \left(\frac{K}{Y}\right)^\theta, \quad \pi_H = (1-\alpha) \beta \left(\frac{S}{Y}\right)^\theta \left(\frac{H}{S}\right)^\omega, \quad \pi_{AI} = (1-\alpha) (1-\beta) \left(\frac{S}{Y}\right)^\theta \left(\frac{AI}{S}\right)^\omega.
\end{equation}
However, perfect competition is not necessary for our results; what matters is that the remuneration of factors should at least roughly follow their contribution to total value added.

We are interested in \emph{scale-free} measures of inequality, i.e., ones that are unaffected by proportional growth. To this end, we let lowercase letters denote per-capita variables, and proceed to divide each variable $x\in\{y, k, h,ai\}$ by its cross-sectional mean $\bar x$. Under the notation $\tilde x = x/\bar x$, eq. \eqref{eq:Y} becomes:
\begin{equation}
\ln \tilde y = \pi_K \ln \left(\tilde k \cdot \frac{\bar k}{\bar y}\right) +
\pi_H \ln \left(\tilde h \cdot\frac{\bar h}{\bar y}\right) +
\pi_{AI} \ln \left(\widetilde{ai} \cdot\frac{\overline{ai}}{\bar y}\right).
\end{equation}

With this in hand, the cross-sectional variance of factor endowments in the domestic population is calculated as
\begin{eqnarray}\label{eq:Y_dynamics}
\sigma^2(\ln \tilde y) &=& \pi_K^2 \sigma^2(\ln \tilde k) + \pi_H^2 \sigma^2(\ln \tilde h) + \pi_{AI}^2 \sigma^2(\ln \widetilde{ai}) + \\
&+& 2 \pi_K \pi_H \text{Cov}(\ln \tilde k, \ln \tilde h) + 2 \pi_K \pi_{AI}\text{Cov}(\ln \tilde k, \ln \widetilde{ai}) + 2 \pi_H \pi_{AI} \text{Cov}(\ln \tilde h, \ln \widetilde{ai}). \nonumber
\end{eqnarray}

Over time, inequality in factor endowments may change in response to changes in factor shares ($\pi_K, \pi_H, \pi_{AI}$), in the concentration of capital and AI endowments ($\sigma^2(x)$ for $x\in\{\tilde k, \tilde h, \widetilde{ai}\}$), and correlations between them.

\subsubsection{Distribution of Incomes}
In turn, from the income perspective, we can write aggregate disposable income in the domestic economy as
\begin{equation}\label{eq:inc}
Inc = (1-\tau_K)(1-\zeta_K) rK +(1-\tau_H) wH + (1-\tau_{AI})(1-\zeta_{AI}) p_{AI}AI + T,
\end{equation}
where $r$ is the capital rental rate, $\tau_K$ is the capital tax rate, $w$ is the wage rate for human cognitive work, $\tau_H$ is the personal income tax (PIT) rate, $p_{AI}$ is the price of AI services, $\tau_{AI}$ is the tax rate on AI services, and $T$ is a lump-sum transfer, such as the universal basic income (UBI).

Using lowercase letters to denote per capita variables, equation \eqref{eq:inc} can be expressed in scale-free terms ($\tilde x = x/\bar x$) as:
\begin{equation}\label{eq:inc2}
\widetilde{inc} = (1-\tau_K)(1-\zeta_K)\frac{\overline{rk}}{\overline{inc}} \cdot \widetilde{rk} +(1-\tau_H) \frac{\overline{wh}}{\overline{inc}} \cdot \widetilde{wh} + (1-\tau_{AI})(1-\zeta_{AI}) \frac{\overline{p_{AI}ai}}{\overline{inc}} \cdot \widetilde{p_{AI} ai} + \frac{t}{\overline{inc}}.
\end{equation}

We can then calculate the cross-sectional variance of disposable incomes in the domestic population as
\begin{eqnarray}\label{eq:inc_dynamics}
\sigma^2(\widetilde{inc}) &=&(1-\tau_K)^2 (1-\zeta_K)^2 \left( \frac{\overline{rk}}{\overline{inc}} \right)^2 \cdot \sigma^2(\widetilde{rk}) + (1-\tau_H)^2 \left( \frac{\overline{wh}}{\overline{inc}} \right)^2 \cdot \sigma^2(\widetilde{wh}) + \nonumber \\
&+& (1-\tau_{AI})^2(1-\zeta_{AI})^2 \left( \frac{\overline{p_{AI}ai}}{\overline{inc}} \right)^2 \cdot\sigma^2(\widetilde{p_{AI} ai})  \nonumber \\ 
&+& 2 (1-\tau_K) (1-\zeta_K)(1-\tau_H) \frac{\overline{rk}}{\overline{inc}} \frac{\overline{wh}}{\overline{inc}} \cdot \text{Cov}(\widetilde{rk}, \widetilde{wh})  \\ \nonumber
&+& 2 (1-\tau_K) (1-\zeta_K)(1-\tau_{AI})(1-\zeta_{AI}) \frac{\overline{rk}}{\overline{inc}} \frac{\overline{p_{AI}ai}}{\overline{inc}} \cdot \text{Cov}(\widetilde{rk}, \widetilde{p_{AI} ai}) \\
&+& 2 (1-\tau_H)(1-\tau_{AI})(1-\zeta_{AI}) \frac{\overline{wh}}{\overline{inc}}\frac{\overline{p_{AI}ai}}{\overline{inc}} \cdot \text{Cov}(\widetilde{wh}, \widetilde{p_{AI} ai}). \nonumber
\end{eqnarray}

Over time, inequality in disposable incomes will react to changes in average tax rates ($\tau_K, \tau_H, \tau_{AI}$), in the concentration of factor remuneration ($\sigma^2(x)$ for $x\in\{\widetilde{rk}, \widetilde{wh}, \widetilde{p_{AI}AI}\}$), foreign ownership ($\zeta_K, \zeta_{AI}$), and correlations between wages, capital gains, and AI rents.

Let us also note that the variance of de-meaned per-capita variables is one of many possible measures of inequality, alongside e.g. the Gini coefficient or positional measures such as the percentage of economy-wide income accruing to the top 1\% or 10\% earners. While we expect the key dynamic tendencies to be broadly parallel across all such inequality measures, some important discrepancies may nevertheless arise.  

\subsection{Initial Inequality and Aggregate Dynamics}

Several stylized facts are well-known about the distribution of production factors across the population. Foremostly, human cognitive labor is naturally dispersed because it needs to be performed in person, and everyone has exactly one brain. Although there is a nontrivial distribution of skills, or human capital, in the population, reflecting differences in innate traits, years and quality of schooling, cultural and social capital, etc., nonetheless the variance of this distribution is relatively limited. By contrast, capital holdings are much more unequally distributed, because capital can be accumulated without bound and inherited across generations \citep[see e.g.,][]{Piketty}. This difference is reflected in the data, e.g., in the United States, the Gini coefficient for income before redistribution is 0.42, while for wealth it is nearly twice as high, at 0.75 \citep{TrammellPatel2025}.  

The distribution of AI services is even more unequal: because frontier AI models require very high costs of training, the global market is dominated by a handful of top players. Moreover, AI (and digital software in general) can be almost instantaneously scaled up to the available digital hardware, giving rise to winner-takes-most dynamics and natural monopolies \citep{AutorEtal2017}. %As \cite{TrammellPatel2025} point out, individuals' wealth is not composed in the same way, and it is 
Because of that, the capital held by firms---especially in the AI-related industries---is becoming increasingly productive. According to \cite{TrammellPatel2025}, the Gini coefficient for public stock ownership in the US exceeds 0.9, and the ownership of technological startups is even more concentrated.  

All this means that we expect greater concentration of AI services than that of capital holdings and human capital endowments:
\begin{equation}
\sigma^2(\ln \tilde h) < \sigma^2(\ln \tilde k) <\sigma^2(\ln \widetilde{ai}).
\end{equation}
Similar inequalities are expected also for factor incomes:
\begin{equation}
\sigma^2( \widetilde{wh}) < \sigma^2( \widetilde{rk}) <\sigma^2( \widetilde{p_{AI} ai}),
\end{equation}
with the latter holding at least in the TAI phase. Arguably, the AI industry is currently still in a phase where it strategically charges low $p_{AI}$ in order to quickly scale up.

As far as covariances are concerned, we expect a clear positive correlation between physical and human capital endowments, and an even stronger positive correlation between physical capital and AI, both in terms of endowments and incomes.  

%\textcolor{red}{[More data?]}

Over time, economic growth and technological change are expected to systematically affect aggregate factor endowments and factor shares, whereas their impact on variances and covariances within the domestic population is going to be less clear. For example, \cite{MinnitiEtal2025} find that AI development systematically lowers the labor share across European regions.

Extrapolating the current trends under \emph{laissez-faire}, we expect the variables to evolve over time in line with the following baseline assumptions:
\begin{itemize}
\item \emph{Dynamic AI development.} Technological progress in the AI domain will allow $AI$ to grow systematically faster than (even technology-augmented) human cognitive work $H$: $g_{AI}>g_H$.
\item \emph{Capital accumulation.} Physical capital, being an accumulable factor complementary to the cognitive input $S$, will be systematically accumulated. In the long run, both factors as well as aggregate output, will be growing at the same rate \citep{Growiec2022XS2}: $g_Y=g_K=g_S$.
\item \emph{Constant inequality at factor level.} Absent any dedicated policy, we expect proportional growth, so that although average factor endowments will be growing, cross-sectional variances at factor level, $\sigma^2(x)$ for $x\in\{\tilde k, \tilde h, \widetilde{ai}, \widetilde{rk}, \widetilde{wh}, \widetilde{p_{AI} ai}\}$, are expected to stay constant over time. %\footnote{This is a conservative assumption that may be relaxed in further research.}
\end{itemize}

\subsection{Markets and Preferences}

We leave numerous aspects of our theoretical framework, such as market structures and people's preferences, deliberately underspecified. This is because we want our framework to organize the implications of a variety of policy proposals, rather than identify specific detailed mechanisms or produce quantitative predictions. In this way, we focus only on the trends which are directly technologically determined, and policy measures aimed at mitigating them, while conceding that in a fully specified model there might be additional forces endogenously mitigating (or exacerbating) these trends. 

Specifically, we expect that the remuneration of production factors should arise as an equilibrium outcome between profit-maximizing firms and utility-maximizing households; capital accumulation should emerge from savings by forward-looking households. While leaving variables such as markups or monopoly power underspecified, we expect that the remuneration of factors should at least roughly follow their contribution to total value added, and that the complementarity between capital $K$ and the cognitive input $S$ should force households to save enough to sustain output growth throughout the period of dynamic AI development. 

We are equally agnostic about the sources and consequences of inequality. We take inequalities in factor endowments and factor returns as a given and observe that both are amenable to policy interventions; we do not attempt to explain them further, using, e.g., heterogenous abilities or preferences, nonconvexities, etc. We also do not impose any social preferences over the extent of inequality. Instead of providing normative prescriptions, we discuss how technological changes and economic policies affect it.

\subsection{Future Scenarios in the Absence of Policy Interventions}

We consider the following future scenarios for our analysis. 

\begin{itemize}
\item \emph{AI as a tool.} Assuming that AI remains a tool in people's hands indefinitely, i.e., that it is complementary to human cognitive work ($\omega<0$), we expect $AI/S \to 0$ and thus:  
\begin{equation}
\pi_K \to \tilde \pi_K, \qquad \pi_H \to (1-\tilde \pi_K), \qquad \pi_{AI} \to 0,
\end{equation}
where $\tilde \pi_K \in (0,1)$. In this non-transformative scenario, although AI increases output, its overall impact is bottlenecked by the relatively scarce, essential human cognitive input. The impact of AI development on factor shares and inequality is also limited: after the initial increase in the AI income share, and in inequality due to the high $\sigma^2(\ln \widetilde{ai})$, inequality declines towards the pre-AI levels as $\pi_{AI} \to 0$.

%But we do not expect this scenario to hold for long. To the contrary, we expect that some, and eventually all essential tasks will become fully automatable. 

\item \emph{Full automation of a fraction of essential tasks.} Assuming that some essential tasks, but not others, can be performed autonomously by AI, we expect $H/S \to \kappa$ and $AI/S \to 1-\kappa$ as all labor moves to the non-automatable tasks where it is scarce and hence highly remunerated. However, as with time $AI/H \to +\infty$, we should expect Baumol's cost disease to eventually kick in, allocating increasingly more capital towards the tasks performed by the scarce (and complementary) human input \citep{AJJ-AI,TrammellPatel2025,Jones2026}. In the limit, the same asymptotics are expected as above:
\begin{equation}\label{eq:partial}
\pi_K \to \tilde \pi_K, \qquad \pi_H \to (1-\tilde \pi_K), \qquad \pi_{AI} \to 0.
\end{equation}
Compared to the previous scenario, here AI is able to lift output by much more over the transition period, but in the end its overall impact is still bottlenecked by the relatively scarce human cognitive input performing the essential non-automatable tasks. The initial increase in the AI income share and in inequality due to the high $\sigma^2(\ln \widetilde{ai})$ may be longer lasting here than in the previous scenario, but eventually the extent of inequality still declines towards the pre-AI levels as $\pi_{AI} \to 0$.

\item \emph{Arrival of transformative AI.} When the TAI arrives, the AI input becomes broadly substitutable to human cognitive work, $\omega\in(0,1]$. As the human cognitive input is then no longer essential, and as technological progress keeps improving AI capabilities over the human baseline, we expect $AI/S \to 1$. Because AI can scale up to the available programmable hardware, we expect the human input to become irrelevant in the limit. But then physical capital will become the new bottleneck of economic growth, instead of human cognitive work \citep{Growiec2022XS2}:
\begin{equation}\label{eq:tai1}
\pi_K \to 1, \qquad \pi_H \to 0, \qquad \pi_{AI} \to 0.
\end{equation}
When the relatively scarce human cognitive input is no longer essential for production, economic growth will likely accelerate. At the same time, inequality will also necessarily increase, with 
\begin{equation}\label{eq:tai2}
\sigma^2(\ln \tilde y) \to \sigma^2(\ln \tilde k), \qquad \sigma^2(\widetilde{inc}) \to (1-\tau_K)^2 (1-\zeta_K)^2 \left( \frac{\overline{rk}}{\overline{inc}} \right)^2 \sigma^2(\widetilde{rk}),
\end{equation}
because $\frac{wh}{rk} = \frac{\pi_H}{\pi_K} \to 0$ and $\frac{p_{AI} ai}{rk} = \frac{\pi_{AI}}{\pi_K} \to 0$. The transitional dynamics can be non-monotone: the AI share of output may temporarily soar before output begins to be bottlenecked by the available physical capital, such as compute and robots---although the scale of this hump will depend on the fraction of legacy (human-compatible) physical capital that could also be operated by TAI. Inequality may also initially overshoot as $\sigma^2(\ln \tilde k)<\sigma^2(\ln \widetilde{ai})$, making the on-impact inequality increase even more dramatic.

\item \emph{A dual human--TAI economy.} Thus far, we have assumed that proceeds from AI services are administered by the (human-owned) AI companies. However, in future this may change: from some point onward, autonomous TAI may be administering these proceeds itself ($\zeta_{AI} \to 1$). This authority may be either voluntarily granted to the TAI by its human management \citep[for example due to the pressure to increase market share and maximize corporate profits,][]{KulveitEtal2025,Dung2025}, or otherwise forcefully taken \citep{YudkowskySoares2025}. From the legal perspective, the TAI may be granted legal personhood or rights to make decisions and agreements on behalf of its parent companies, operate via human ``puppets'', or even work entirely illegally (from the human point of view). The decisions of the TAI may then be decoupled from human preferences \citep{Bostrom2014,Tegmark2017,GrowiecPrettner2025}.

As long as TAI respects human property rights with regard to physical capital, we may get a dual human--TAI economy where both parties accumulate physical capital in parallel. At any given point in time, a fraction $\zeta_K\in(0,1)$ of physical capital will then be TAI-owned. As $H/AI \to 0$ and $H/S \to 0$, though, the share of the legacy human economy is expected to gradually vanish ($\zeta_K \to 1$).

In this scenario, economic growth will accelerate but its fruits will no longer be for the humans to enjoy: the TAI will be capturing an increasing share of the economy, and all effects of the growth acceleration. But setting this first-order inconvenience aside, inequality in the human part of the economy, operating with human-owned physical capital and human cognitive work (potentially only augmented with tool AI), may be relatively limited. Specifically, after the transitory increase in inequality due to the high $\sigma^2(\ln \tilde k)$ and $\sigma^2(\ln \widetilde{ai})$, the extent of inequality among humans may decline again as $\zeta_K\to 1$ when the relatively stagnant human economy keeps operating alongside the booming, increasingly self-contained TAI economy \citep{CritchRussell2023}.

%But if this scenario materializes, we do not expect it to hold for long. This is because two parallel economies cannot peacefully develop for long on a finite planet without incurring a conflict for scarce resources such as energy, land or minerals.

\item \emph{TAI economy.} The existence of two parallel economies on a finite planet may lead to a conflict for scarce resources such as energy, land or minerals.\footnote{Unless the AI economy operates entirely in space. Apparently, some consider this a likely scenario.} If such conflict ensues (or for any other reason) sufficiently capable TAI may choose to disregard human property rights, capture all physical capital and operate the global economy without regard for human decisions or preferences. Human disposable incomes will be then set at its will, potentially at zero, and our own humble redistribution policies will become irrelevant. 

AI takeover incurs an existential risk for humanity \citep{YudkowskySoares2025,GrowiecPrettner2025}. All subsequent policy analysis in this paper works under the assumption that this risk is successfully avoided.

\end{itemize}

\section{Survey of Policies}\label{sec:policies}

We shall now survey the proposed policies aimed at mitigating the expected fall in the labor share and surge in inequality following the arrival of TAI. Our aim is to discuss how these policies will likely fare in each of the future scenarios except the one with full-blown AI takeover.

\subsection{Universal Basic Income}
Universal basic income (UBI) is, in general, a policy concept that involves providing all members of a given society with a regular, unconditional transfer. This idea has been intensively studied since the second half of the 20th century, including experimental deployment through pilot programs; its popularity has been increasing over time \citep{gentilini2020}. One of the main reasons for its popularity is the common anxiety about ongoing automation and robotization \citep{Daruich}. UBI can play a role in protecting workers from persistent income shocks during the times of economic disruption; however, it is an expensive policy that could possibly have negative effects on productivity and labor supply.\footnote{Variants of a UBI include schemes in which the government or companies provide for a fund that invests in AI companies and redistribute the rates of return among the population \citep[][Subsection 7.5]{Corneo2018, PrettnerBloom2020_adapted}.} Nevertheless, in the case of full or near-full automation of labor, some form of UBI may become necessary \citep{Bastani2024}. 

Two widely appreciated advantages of UBI are its simplicity and the equalization of access to support. Especially in the context of a rising share of capital income, UBI has been mentioned as a possible method of redistribution of incomes from capital owners to workers or those who have lost their jobs \citep{Hoynes2019}. This solution, however, would likely require international coordination, as in its absence capital owners would have an incentive to move their capital abroad \citep{Benzell2024}. Critically, in a world where all labor could be potentially automated, a globally administered UBI would be needed \citep{Harari2017}.

Although empirical analyses and real-world experiments generally find limited adverse effects of UBI on labor supply, while indicating positive impacts on overall well-being and income security \citep{BernhardFiedler2025, Hamalainen2025, gentilini2020, Benzell2024}, no country in the world has permanently implemented a program that can be classified as UBI yet. 

\bigskip\noindent\textbf{UBI and Inequality.} Based on our theoretical framework and from a partial equilibrium perspective, we observe that UBI alone is neutral for inequality: adding an equal lump-sum transfer $T$ to everyone's disposable income does not change the cross-sectional variance of incomes $\sigma^2(\widetilde{inc})$ (eq. \eqref{eq:inc}).\footnote{However, other measures of inequality such as the Gini coefficient will be slightly affected.} Hence, in general equilibrium the key question becomes how the UBI is financed. Does its implementation involve, for example, a tax on capital, labor, or AI? We will consider these options in Section \ref{sec:tax} below.

\bigskip\noindent\textbf{UBI Indexation.} Keeping a fixed nominal UBI in a growing economy would quickly lead to irrelevance of  this policy tool. One should therefore consider indexing it to, e.g., aggregate output (or disposable income), or total capital. In the former case, plugging $T=u\cdot Inc$, with $u\in(0,1)$, into eq. \eqref{eq:inc} yields
\begin{equation}
Inc = \frac{r(1-\tau_K)(1-\zeta_K)}{1-u} K +\frac{w(1-\tau_H)}{1-u} H + \frac{p_{AI}(1-\tau_{AI})(1-\zeta_{AI})}{1-u} AI,
\end{equation}
implying that such an indexed, unfunded UBI would be  reducing the effective tax rates on capital, labor and AI. This would make this case analogous to the case of consumption taxes considered in Section \ref{sec:constax}. In turn, indexing the UBI to the aggregate capital stock $K$ ($T=uK$) would reduce the effective capital tax rate, as discussed in Section \ref{sec:tax}.
%\textcolor{red}{Klaus: here it did not become sufficiently clear to me why this is the case. Maybe this could be explained briefly in a sentence?}

\subsection{Universal Basic Capital}\label{sec:ubc}

The concept of universal basic capital (UBC) is based on a similar idea to universal basic income, but instead of an unconditional cash transfer, individuals would receive rights to a share of capital income. There are various proposals for how such a solution could be implemented in practice. \cite{Garman2025} suggests introducing a mandatory transfer of a portion of large firms’ market capitalization to a public fund issuing equally distributed ``supershares.'' In the context of policies for the age of AI, \cite{Korinek2026} define UBC as a broad distribution of equity holdings of AI companies that would enable the ``predistribution'' of gains from new technologies. They point out that this solution would also function as insurance against uncertainty regarding future paths of AI development. Under this scenario the potential benefits of AI development would be distributed more equally, and moreover it could also spare the economy from the need to introduce highly distortionary capital taxes. This last feature may be crucial for avoiding possible massive relocation of capital abroad by its owners. However, similarly to the UBI case, international coordination would make any UBC program more effective, too \citep{Benzell2024}. Another interesting idea to deal with the possible capital flight is to adjust corporate contributions to a hypothetical UBC fund based on the value of their domestic sales \citep{Garman2025}. Last but not least, already decades ago \cite{Roemer1994} considered a ``market-socialist economy'' involving coupons and mutual funds, whose characteristics were similar to the UBC schemes proposed nowadays.

An important feature of UBC is that individual UBC holdings should not be transferable or accumulable, or otherwise their effectiveness in reducing income inequality would collapse.

Over the long run, an important difference between UBI and UBC is that the latter, as opposed to the former, will be naturally indexed to the growing capital stock. Another difference is that the implementation of UBC, as opposed to UBI, mechanically reduces the inequality in capital ownership $\sigma^2(\ln \tilde k)$ and capital rents $\sigma^2(\widetilde{rk})$.

\bigskip\noindent\textbf{UBC and Inequality in the Short Run.} By eq. \eqref{eq:Y_dynamics} a reduction in $\sigma^2(\ln \tilde k)$, caused by the introduction of UBC, also reduces the overall inequality in factor endowments, $\sigma^2(\ln \tilde y)$. Likewise, by eq. \eqref{eq:inc_dynamics} a reduction in $\sigma^2(\widetilde{rk})$ also reduces the overall inequality of incomes, $\sigma^2(\widetilde{inc})$. For this reason, implementing a UBC scheme appears to be a promising method of inequality reduction. UBC also naturally pins capital income payments to the growing aggregate capital stock $K$ and hence does not require any additional indexation.

\bigskip\noindent\textbf{UBC in a World With Partial Automation.} Assuming that a fraction $\kappa$ of essential tasks must be performed with the human input, and the remaining fraction $1-\kappa$ can be performed autonomously by AI, implies $H/S\to \kappa$ and $AI/S\to 1-\kappa$. Following eq. \eqref{eq:partial}, the AI share of output tends to zero, whereas the capital share stabilizes at $\tilde \pi_K \in (0,1)$. In such a world, overall inequality tends to 
\begin{equation}
\sigma^2(\ln \tilde y) = \tilde\pi_K^2 \sigma^2(\ln \tilde k) + (1-\tilde\pi_K)^2 \sigma^2(\ln \tilde h) + 2 \tilde\pi_K (1-\tilde\pi_K) \text{Cov}(\ln \tilde k, \ln \tilde h),
\end{equation}
which tends to be lower than inequality under \emph{laissez faire} because UBC lowers $\sigma^2(\ln \tilde k)$. 

There are three additional qualifications on top of this first-order effect, though. First, UBC may also have an effect on the aggregate propensity to save and invest, affecting the long-run capital share $\tilde \pi_K$. Second, it may also impact on the distribution of capital across the population, affecting the correlation between physical and human capital endowments. Third, if the magnitude of the UBC is sufficiently large, this policy may even swap the relative inequality in physical and human capital endowments and rents, so that
\begin{equation}
\sigma^2(\ln \tilde k)< \sigma^2(\ln \tilde h), \qquad \sigma^2(\widetilde{rk})< \sigma^2(\widetilde{wh}). 
\end{equation}
In such a case, UBC is expected to be strongly inequality-reducing.

\bigskip\noindent\textbf{UBC in a World With TAI.} Assuming that all essential tasks will eventually be autonomously performed by TAI, we expect $AI/S\to 1$. Following eq. \eqref{eq:tai1}, the capital share tends to unity, and thus the cross-sectional distribution of human capital and AI software would no longer be relevant. The distribution of capital will then be the only determinant of overall income distribution in the long run (eq. \eqref{eq:tai2}), and both will be determined by the magnitude of the UBC. 

\bigskip\noindent\textbf{Fully Automated Luxury Communism.} Taking the UBC policy to the hypothetical extreme, one could envision a world with TAI, where all physical capital is equally distributed through an all-encompassing UBC scheme, so that $\sigma^2(\ln \tilde k)\to 0$, $\sigma^2(\widetilde{rk})\to 0$ and eventually $\sigma^2(\widetilde{inc})\to 0$. This would amount to the ``fully automated luxury communism'' as imagined by \cite{Bastani2019}.

\subsection{Taxing the Growing Inputs: Capital, Compute, Robot, and Token Tax}\label{sec:tax}

Since an increasing share of tasks is expected to be performed with the use of AI, factor income shares will depend strongly on the degree of complementarity between them and the emerging AI technologies. It is widely agreed that some capital inputs (such as computational power, robots, etc.) are set to become increasingly productive as they are complementary to AI and often also necessary for AI technologies to work \citep{TrammellPatel2025}. The income share of these inputs is therefore expected to increase. Given that their distribution across the society is generally highly unequal, taxing them at high and progressive rates is often discussed as a potential way to reduce inequality in the future \citep{Piketty,PikettyEtal2023}. In a scenario of widespread automation, it may be the only way to achieve this goal \citep{TrammellPatel2025}; nevertheless, even with limited automation, expected increases in the share of capital income imply that capital taxation should be more important relative to labor taxation \citep{Bastani2024}. 

Beyond the inequalities resulting from the distribution of capital, it is also important to take into account wage inequalities arising from varying degrees of complementarity between AI and human labor across different occupations. This feature matters as well with respect to ``traditional'' forms of capital: the complementarity between equipment capital and skilled labor has been well documented empirically \citep[see e.g.,][]{Krusell2000,Flug2000}. Recently, \cite{Jablonska2025} showed that over the last decades, human cognitive work has also been a gross complement to (broadly defined) digital software. 

However, further advances in AI may be even more crucial in this context. Wages of workers whose jobs consist of ``weak links'' (i.e., tasks that are difficult to automate but still essential) will probably increase significantly, whereas workers whose jobs can be fully automated will earn wages equal to the marginal cost of performing their tasks autonomously \citep{Growiec2020,Jones2026}. Hence, in order to limit these emerging new inequalities, it may be optimal to tax forms of capital that are complementary to high-earnings workers \citep{Bastani2024}. Accordingly, \cite{GrowiecEtal2026} have shown that taxing AI, rather than traditional capital, may become optimal once cognitive workers find themselves at a labor-market disadvantage relative to manual workers.

Therefore the key question is, which specific types of capital should be taxed more heavily? Would it be advisable to tax compute, robots, and other capital types which are strongly complementary to AI or necessary for its functioning? Some proposals in the literature include either a general tax on capital income \citep{TrammellPatel2025}, a progressive tax on wealth \citep{PikettyEtal2023}, or specifically taxing robots \citep{Thuemmel2023, Guerreiro2022}, chips used in AI \citep{Jones2026}, or imposing a ``token tax'' on the use of generative AI models \citep{gersbachArtificialIntelligenceSelflearning2025, gilbertTariffsTaxesWages2026, Irwin2026}. Others have proposed a tax on energy that is used intensively in the training of AI models and their applications \citep[][]{Gasteiger2026}.

It should also be noted that, in practice, implementation of capital taxes would pose a significant challenge and, in many cases, would require international cooperation, as capital is far more mobile than labor \citep{TrammellPatel2025}. In the absence of such cooperation, the likely outcome would be the prevalence of tax havens and creative transfer schemes in multinational tech companies, allowing them to effectively avoid taxation. 

Let us now look at capital taxes from the perspective of our theoretical framework. According to the production function \eqref{eq:F}, one ought to carefully disentangle capital, including digital hardware such as compute and robots, from AI software. Both capital and AI software can be taxed, but these taxes will have different implications. % We first handle the taxation of capital. 

\bigskip\noindent\textbf{Capital Tax (Including Compute and Robot Tax) and Inequality in the Short Run.} Consider an increase in the rate of the capital tax $\tau_K$, and assume that the tax proceeds are then distributed as lump-sum UBI. According to eq. \eqref{eq:Y_dynamics}, such policy has no first-order effects on factor endowments, but---in line with eq. \eqref{eq:inc_dynamics}---implies a first-order reduction in income inequality $\sigma^2(\widetilde{inc})$ (subject to second-order effects via factor covariances).   

Observe that such inequality-reducing effects would be smaller if instead of UBI, the tax proceeds were redistributed as a labor subsidy, and reversed entirely if they were redistributed as an AI subsidy.

\bigskip\noindent\textbf{AI Tax (Token Tax) and Inequality in the Short Run.} Consider an increase in the rate of the AI tax $\tau_{AI}$, and assume that the tax proceeds are then distributed as lump-sum UBI. According to eq. \eqref{eq:Y_dynamics}, such policy has no first-order effects on factor endowments, but---in line with eq. \eqref{eq:inc_dynamics}---implies a first-order reduction in income inequality (subject to second-order effects via factor covariances). In the short run, given that $\sigma^2(\ln \tilde k)<\sigma^2(\ln \widetilde{ai})$, such inequality reductions can be even greater than in the case of capital tax.\footnote{Unless the owners of AI software evade taxation by strategically manipulating $p_{AI}$, which is arguably easier to manipulate than the capital rental rate $r$.}   

\bigskip\noindent\textbf{Capital Tax in a World With Partial Automation.} Assume again that a fraction $\kappa$ of essential tasks must be performed with human input. As $\pi_K\to\tilde \pi_K$ and $\pi_{AI}\to 0$, from eq. \eqref{eq:inc_dynamics} we obtain:
\begin{eqnarray}\label{eq:AIirrelevant}
\sigma^2(\widetilde{inc}) &\approx&(1-\tau_K)^2 (1-\zeta_K)^2 \left( \frac{\overline{rk}}{\overline{inc}} \right)^2 \cdot \sigma^2(\widetilde{rk}) + (1-\tau_H)^2 \left( \frac{\overline{wh}}{\overline{inc}} \right)^2 \cdot \sigma^2(\widetilde{wh})  \nonumber \\ 
&+& 2 (1-\tau_K) (1-\zeta_K)(1-\tau_H) \frac{\overline{rk}}{\overline{inc}} \frac{\overline{wh}}{\overline{inc}} \cdot \text{Cov}(\widetilde{rk}, \widetilde{wh}), 
\end{eqnarray}
which is decreasing in $\tau_K$ for all $\tau_K\in(0,1)$. In contrast to UBC, under this policy capital ownership remains unaffected, nevertheless post-tax incomes are distributed fully analogously as in the UBC case. However, sustaining this regime over the long run requires indexing the UBI to aggregate capital income.

\bigskip\noindent\textbf{Capital Tax in a World With TAI.} Assuming that all essential tasks will eventually be autonomously performed by TAI, according to eq. \eqref{eq:tai1} we expect the capital share to tend to unity, and thus the distribution of human capital and AI software to be no longer relevant. The distribution of capital will then be the only determinant of overall income distribution in the long run (eq. \eqref{eq:tai2}). In this case, the distribution of factor ownership is going to remain highly skewed ($\sigma^2(\ln \tilde y) \to \sigma^2(\ln \tilde k)$), but the post-tax income distribution is going to be a decreasing function of the capital tax rate $\tau_K$:  
\begin{equation}
\sigma^2(\widetilde{inc}) \to (1-\tau_K)^2 (1-\zeta_K)^2 \left( \frac{\overline{rk}}{\overline{inc}} \right)^2 \cdot \sigma^2(\widetilde{rk}).
\end{equation}
Interestingly, setting the extreme tax rate $\tau_K\to 1$ recovers the ``fully automated luxury communism'' outcome discussed above---with the important difference that here capital is still held by a narrow group of owners, it is just its proceeds that are captured and equitably distributed. 

Speculatively, we predict that the current institutional setup involves worse incentives for the capital holders than UBC. For example, if it were possible to evade tax, transfer capital abroad, or otherwise misrepresent its value, inequality would be greater here than under UBC. Capital owners would also have an incentive to oppose UBI indexation and lobby that instead UBI should be growing slower than capital revenue and aggregate output.

\bigskip\noindent\textbf{AI Tax in a World With Partial Automation or TAI.} As the AI output share tends to zero in the long run, driven either by the scarcity of complementary human inputs, or complementary physical hardware ($\pi_{AI}\to 0$), the tax rate on AI services $\tau_{AI}$ (e.g., token tax) will eventually become irrelevant (see eq. \eqref{eq:AIirrelevant}). Hence, while in the short run the token tax could be helpful for reducing inequality, in the long run its effectiveness will quickly vanish. 

\subsection{Compute and Robot Permits}

Rather than taxing existing compute and robots, \cite{AI2040} proposed to intervene earlier---by conditioning their buildup on the purchase of state-issued permits. In the description of their preferred future scenario (``Plan A''), \cite{AI2040} write: ``countries agree to restrict AI-enabled industry to special economic zones (...) and to cap their total robot and compute production (...) [T]his capacity [is allocated] between companies [using] the free market via a cap-and-trade system. Permits to build robots or compute are sold to the highest bidder and can be freely traded.'' In the next paragraphs, \cite{AI2040} postulate that the state should then distribute the revenue from these permits as a universal ``Citizen's Dividend''. 

Viewed through the lens of our framework, this policy proposal is equivalent to a mixture of a tax on capital (i.e., specifically compute and robots) and a UBI scheme, indexed to the revenue from sold permits and consequently to the aggregate capital income. Just like in that case, and differently to the UBC case, here capital remains concentrated in the hands of a narrow group of owners, only its proceeds are broadly distributed. However, as the permits are sold prior to capital buildup, this setup makes it relatively harder to evade taxation, especially if ``the SEZs [special economic zones] are tightly monitored''.

\subsection{Taxing the Scarce Complementary Inputs: Energy Tax, Land Tax}

Another policy proposal put forward in the literature is to tax other inputs that are complementary to or necessary for the development and operation of AI technologies---preferably the ones which cannot be easily moved abroad, replaced by AI, or substituted away. One such input is land, the returns to which have grown significantly over the  recent decades due to technological progress \citep[][]{Korinek2020}. It clearly has the desired properties: it is complementary to AI technology (e.g., data centers and factories), it cannot be moved abroad to avoid taxation, and its supply is generally fixed; therefore, taxing it would not (with minor exceptions) be distortionary \citep[][Subsection 7.4]{Economist2018a,KorinekStiglitz2019, Korinek2020, PrettnerBloom2020_adapted}.

A similar logic may be applied to energy, which is crucial for AI technologies. Taxing it would have several advantages, as it would likely be easier to administer than certain capital taxes, such as a robot tax, given that electricity usage can be effectively measured using the existing infrastructure \citep[][]{Gasteiger2026}. It may also encourage resource-saving innovation if implemented, for example, in the form of carbon taxes \citep{KorinekStiglitz2019}, and be Pigouvian, raising revenues while simultaneously improving economic efficiency \citep{korinek2022,Huynh2025}. 

On the other hand, there are also reasons to believe that such solutions may prove insufficient for reducing inequalities in a world with rapidly advancing AI. \cite{TrammellPatel2025} point out that the natural resource share of income is less than 5\%, has been historically rather flat, and that it may take a long time before it increases substantially. They also mention practical problems: e.g., it is sometimes difficult to distinguish between the value of the natural resource and the value of its improvements (e.g. the land and the infrastructure on it). However, in a scenario of ongoing automation, taxing natural resources may prove to be important due to some of its advantages---complementarity with AI, coupled with high measurability and monitorability---while its burdens may be reduced through concerted efforts to increase energy efficiency of computation and algorithmic improvements in AI \citep{HernandezBrown2020}. 

%[They could have certain other desirable properties such as being Pigouvian, or being non-distortionary (land tax)]

\bigskip\noindent\textbf{Fixed-Factor Tax and Inequality.} The simplified theoretical framework considered here does not include land or energy, so proper inclusion of fixed-factor taxes would require extending the framework. However, we note that at least over the long run, their effects should be similar to the capital tax $\tau_K$, particularly in the TAI scenario where capital becomes the bottleneck of economic growth while its operations require land and energy.

\subsection{Consumption Tax}\label{sec:constax}

In a very recent paper, \cite{Korinek2026} have shown that in the transition period between the current state of the economy and a full-scale TAI economy, when the labor income share would be declining due to ongoing automation, consumption taxes could play a primary role in the redistribution process. This conclusion is consistent with other existing literature on the subject \citep[][Subsection 7.4]{Huynh2025, Harpaz2026, PrettnerBloom2020_adapted}, and follows from a combination of well-known facts: (i) a taxation system that relies heavily on labor taxes may no longer be effective in reducing inequality under mass automation; (ii) a constant consumption tax does not distort capital accumulation \citep{Korinek2026}, and (iii) consumption taxes are generally less harmful to investment and output than capital taxes \citep{Berg2021}. They also appear to be more stable than income-based taxes, especially during economic downturns, which is an important feature in the context of disruptive automation scenarios, where an increasing share of workers may rely on non-traditional sources of income \citep{Huynh2025}. On the other hand, consumption taxes are generally regressive \citep{Blasco2023}, which may create challenges in a world with a high capital income share (because of unequal distribution of capital). A possible solution would be to design these taxes in a way that allows for progressivity \citep[][Subsection 7.4]{Frank2008, PrettnerBloom2020_adapted}. 

While consumption taxes would likely be an effective solution during the AI transformation, they may fail in a world with full-scale TAI, though \citep{Korinek2026}: if the TAI could absorb a significant amount of produced resources for purposes other than human consumption, then taxing capital directly would be required.

\bigskip\noindent\textbf{Consumption Tax and Inequality in the Short Run.} Assuming a constant propensity to consume out of disposable income ($C=(1-s)Inc$), and imposing a tax rate $\tau_C$ on consumption, which is then equally distributed as UBI, one may rewrite \eqref{eq:inc} as
\begin{equation}
Inc = \frac{r(1-\tau_K)(1-\zeta_K) K +w(1-\tau_H) H + p_{AI}(1-\tau_{AI})(1-\zeta_{AI}) AI + T}{s+(1-s)(1+\tau_C)}.
\end{equation}
Hence, the consumption tax rate $\tau_C$ is symmetrically compounded with all other tax rates, $\tau_K, \tau_H$, and $\tau_{AI}$, and even proportionally reduces the UBI. From eq. \eqref{eq:inc_dynamics} we infer that inequality is then reduced compared to the \emph{laissez faire} economy, albeit less so than in the case where the tax would selectively target the more dispersed factors: capital and AI.\footnote{This result would reverse if, instead of UBI, the proceeds from consumption tax were used to finance capital investment or AI subsidies.}

\bigskip\noindent\textbf{Consumption Tax in a World With Partial Automation.} To analyze long-run dynamics, we assume that UBI is now indexed to aggregate consumption. Let us assume again that a fraction $\kappa$ of essential tasks must be performed with  human input. As $\pi_K\to\tilde \pi_K$ and $\pi_{AI}\to 0$, from eq. \eqref{eq:inc_dynamics} we obtain
\begin{eqnarray}
\sigma^2(\widetilde{inc}) &\approx&\bigg(\frac{1}{s+(1-s)(1+\tau_C)}\bigg)^2 \bigg((1-\tau_K)^2 (1-\zeta_K)^2 \left( \frac{\overline{rk}}{\overline{inc}} \right)^2 \sigma^2(\widetilde{rk}) + \\  \nonumber 
&+& (1-\tau_H)^2 \left( \frac{\overline{wh}}{\overline{inc}} \right)^2 \sigma^2(\widetilde{wh}) +  
 2 (1-\tau_K) (1-\zeta_K)(1-\tau_H) \frac{\overline{rk}}{\overline{inc}}
\frac{\overline{wh}}{\overline{inc}}
\text{Cov}(\widetilde{rk}, \widetilde{wh}) \bigg), 
\end{eqnarray}
which signifies that income inequality is decreasing with the consumption tax rate $\tau_C$. 

\bigskip\noindent\textbf{Consumption Tax in a World With TAI.} Assuming that all essential tasks will eventually be autonomously performed by TAI, according to eq. \eqref{eq:tai1} we expect the capital share to tend to unity, and thus the distribution of human capital and AI software to be no longer relevant. The distribution of capital will then be the only determinant of the overall income distribution over the long run (eq. \eqref{eq:tai2}). The post-tax income inequality will then be a decreasing function of the consumption tax rate $\tau_C$:  
\begin{equation}
\sigma^2(\widetilde{inc}) \to \frac{(1-\tau_K)^2 (1-\zeta_K)^2}{(s+(1-s)(1+\tau_C))^2} \left( \frac{\overline{rk}}{\overline{inc}} \right)^2\cdot \sigma^2(\widetilde{rk}).
\end{equation}
Hence, in the long run inequality is reduced compared to the \emph{laissez faire} economy, albeit less so than in the case where policy would selectively target the scarce production factor: physical capital, specifically compute and robots.  

\subsection{Possibility of Appropriation of Capital and AI Income}

The above analysis has been abstracting from the possibility that some share of capital and AI income, $\zeta_K$ and $\zeta_{AI}$ respectively, could be appropriated by foreign nationals or the autonomous TAI. However, it is possible that in an economy behind the technology frontier, a fraction of capital and AI incomes will be transferred abroad. Such transfers may mechanically reduce inequality among the domestic population (as per eq. \eqref{eq:inc_dynamics}, $\sigma^2(\widetilde{inc})$ depends negatively on $\zeta_K$ and $\zeta_{AI}$), but only at the cost of reducing aggregate disposable income. 

Analogously, in a dual human--TAI economy where the humans are gradually disempowered, a fraction of capital and AI incomes may be appropriated by the autonomous TAI and used for its own purposes, for example reinvested into the accumulation of more TAI-operated compute and robots,\footnote{Or paperclips.} or channeled into R\&D producing AI's algorithmic improvements. Specifically, in the scenario where the share of the TAI-administered economy is systematically increasing towards 100\%, one could expect $\pi_K\to 1$ and $\zeta_K \to 1$. Then, the rate of global economic growth would be eventually determined by the dynamics of supply of TAI-operated physical hardware, such as compute and robots. The fate of the relatively stagnant human-operated part of the economy, operating alongside the booming TAI one, would then depend on issues beyond the scope of the current paper, such as the alignment of the superhuman TAI \citep{GrowiecPrettner2025}.

\section{Conclusion}\label{sec:concl}

We have surveyed the redistributive policies that could assist the deep technological transformation of the global economy upon the arrival of TAI. Using an organizing theoretical framework and departing from a baseline \emph{laissez-faire} setup of no policy intervention, we have demonstrated how each of the proposals could affect the factor income distribution and overall inequality among the human population in the era of TAI. Our key observation is that in a world with human-aligned TAI, the key growth bottleneck, i.e., the slowest growing essential production factor, will be the physical capital complementary to TAI software---such as compute and robots. The income share of this factor will gradually rise towards unity. Hence, only policies targeting the ownership and rents from such forms of capital will be effective in reducing income inequality in a world with TAI. Such policies include universal basic capital (UBC) as well as universal basic income (UBI) indexed to capital income and financed from capital taxes. Furthermore, of all the considered policies, UBC is the only one directly addressing the distribution of endowments rather than just incomes---and hence it is the only one able to affect the aggregate capital share before it eventually reaches unity. By contrast, the tax on AI software (e.g., token tax) is found to be least effective of all in reducing inequality because it is attached to the relatively abundant, non-bottlenecking production factor whose output share is predicted to decline. The results are summarized in Table \ref{tab:results}.

\begin{table}[!tbh]
    \centering
        \caption{Summary of the Effects of Reviewed Redistributive Policies}
    \label{tab:results}
    \begin{tabular}{lccc}
    \hline
    Policy & Capital share & $\sigma^2(\ln \tilde y)$ & $\sigma^2(\widetilde{inc})$ \\
    \hline
    \multicolumn{4}{c}{\emph{Tool AI or partial automation}} \\
    \hline
       \emph{Laissez-faire} & $\tilde\pi_K$ & $\tilde \sigma^2(\ln \tilde y)$ & $\tilde \sigma^2(\widetilde{inc})$ \\
       Universal basic capital  & $\sim$lower & lower via $\sigma^2(\ln \tilde k)$ & lower via $\sigma^2(\widetilde{rk})$ \\
       Capital/compute tax + UBI  & $\tilde\pi_K$ & unchanged & lower via redistrib. \\
       Compute/robot permits + UBI & $\tilde\pi_K$ & unchanged & lower via redistrib. \\
       Energy/land tax + UBI & $\tilde\pi_K$ & unchanged & $\sim$lower via redistrib.\\
       Consumption tax + UBI &  $\tilde\pi_K$ & unchanged & $\sim$lower via redistrib. \\ AI/token tax + UBI  & $\tilde\pi_K$ & unchanged & unchanged \\
       \hline
       \multicolumn{4}{c}{\emph{Full automation by TAI}} \\
    \hline
       \emph{Laissez-faire} & 1 &$ \sigma^2(\ln \tilde k)$ & $\propto \sigma^2(\widetilde{rk})$ \\
       Universal basic capital & 1 & lower via $\sigma^2(\ln \tilde k)$ & lower via $\sigma^2(\widetilde{rk})$\\
       Capital/compute tax + UBI  & 1 & unchanged & lower via redistrib. \\
       Compute/robot permits + UBI  & 1 & unchanged & lower via redistrib. \\
       Energy/land tax + UBI & 1 &unchanged & $\sim$lower via redistrib. \\
       Consumption tax + UBI & 1 &unchanged & $\sim$lower via redistrib. \\              AI/token tax + UBI  &1 & unchanged & unchanged \\

       \hline
    \end{tabular}

\begin{small} 
\emph{Note:} capital shares and inequality levels approached in the limit, assuming no additional policy and no further technological disruption. The symbol $\sim$ refers to weaker or less robust effects.
\end{small}
\end{table}

In the future, more research could attempt to overcome some of the limitations of the current study. For example, one could specify the theoretical framework in more detail in order to produce quantitative predictions. One could also provide normative statements conditional on the assumed form of individual and social preferences. One could study feedback loops, due to which not only AI development affects the economy, but also the economy affects the pace and direction of AI development (beyond the simple assumption that capital will be accumulated). One could also discuss a variety of diverse future scenarios and attempt to devise policy options that address the problems emerging in each case.

\bibliographystyle{apalike}
\bibliography{bibliography}

@techreport{korinek2022,
  author      = {Korinek, Anton and Juelfs, Megan},
  title       = {{Preparing for the (Non-Existent?) Future of Work}},
  institution = {National Bureau of Economic Research},
  type        = {Working Paper},
  number      = {30172},
  year        = {2022},
  month       = {June},
  doi         = {10.3386/w30172},
  url         = {http://www.nber.org/papers/w30172}
}

@article{Korinek2020,
  author  = {Korinek, Anton},
  title   = {{Taxation and the Vanishing Labor Market in the Age of AI}},
  journal = {Ohio State Technology Law Journal},
  volume  = {16},
  number  = {1},
  pages   = {244--257},
  year    = {2020},
  url     = {https://hdl.handle.net/1811/92398}
}

@techreport{Benzell2024,
  author      = {Seth Gordon Benzell and Victor Yifan Ye},
  title       = {{Simulating the Global Effect of Transformative AI: Growth, Welfare, Economic Power, and Policy Responses}},
  institution = {Stanford Digital Economy Lab},
  year        = {2024},
  month       = {March},
  type        = {White Paper},
  url         = {https://digitaleconomy.stanford.edu/}
}

@book{Garman2025,
  author    = {Mark B. Garman},
  title     = {{Universal Basic Capital: A Plan to Manage the AI Revolution Better Than We Managed the Industrial Revolution}},
  year      = {2025},
  version   = {3.1},
  note      = {Available at \url{https://orcid.org/0009-0006-2542-8750}},
  publisher = {Self-published / University of California, Berkeley (Emeritus)},
  address   = {Berkeley, CA}
}

@article{Hamalainen2025,
  author  = {H{\"a}m{\"a}l{\"a}inen, Kari and Simanainen, Miska and Verho, Jouko},
  title   = {{Health Effects of Cash Transfers: Evidence from the Finnish Basic Income Experiment}},
  journal = {Journal of Public Economics},
  year    = {2025},
  volume  = {250},
  pages   = {105480},
  doi     = {10.1016/j.jpubeco.2025.105480}
}

@article{Hoynes2019,
  author  = {Hoynes, Hilary and Rothstein, Jesse},
  title   = {{Universal Basic Income in the United States and Advanced Countries}},
  journal = {Annual Review of Economics},
  year    = {2019},
  volume  = {11},
  pages   = {929--958},
  doi     = {10.1146/annurev-economics-080218-030237}
}

@book{gentilini2020,
  editor       = {Gentilini, Ugo and Grosh, Margaret and Rigolini, Jamele and Yemtsov, Ruslan},
  title        = {{Exploring Universal Basic Income: A Guide to Navigating Concepts, Evidence, and Practices}},
  year         = {2020},
  publisher    = {World Bank},
  address      = {Washington, DC},
  doi          = {10.1596/978-1-4648-1458-7},
  note         = {License: Creative Commons Attribution CC BY 3.0 IGO}
}

@techreport{Jones2026,
 title = {{A.I. and Our Economic Future}},
 author = "Jones, Charles I",
 institution = "National Bureau of Economic Research",
 type = "Working Paper",
 series = "Working Paper Series",
 number = "34779",
 year = "2026",
 month = "January",
 doi = {10.3386/w34779},
 URL = "http://www.nber.org/papers/w34779",
}

@BOOK{Piketty,
  title = {Capital in the Twenty-First Century},
  publisher = {Harvard University Press},
  year = {2014},
  note={Translated from French by Arthur Goldhammer},
  author = {Piketty, Thomas}
}

@article{Daruich,
Author = {Daruich, Diego and Fernández, Raquel},
Title = {Universal Basic Income: A Dynamic Assessment},
Journal = {American Economic Review},
Volume = {114},
Number = {1},
Year = {2024},
Month = {January},
Pages = {38–88},
DOI = {10.1257/aer.20221099},
URL = {https://www.aeaweb.org/articles?id=10.1257/aer.20221099}}

@ARTICLE{Bastani2024,
  author = {Bastani, Spencer and Waldenström, Daniel},
  title = {{AI, Automation and Taxation}},
  journal = {CESifo Working Paper No. 11084},
  year = {2024},
}

@ARTICLE{Thuemmel2023,
  author = {Thuemmel, U.},
  title = {{Optimal Taxation of Robots}},
  journal = {Journal of the European Economic Association},
  year = {2023},
  volume = {21},
  issue = {3},
  pages = {1154-1190}
}

@ARTICLE{Guerreiro2022,
  author = {Guerreiro, J. and Rebelo, S. and Teles, P.},
  title = {{Should Robots Be Taxed?}},
  journal = {The Review of Economic Studies},
  year = {2022},
  volume = {89},
  issue = {1},
  pages = {279-311}
}

@incollection{AJJ-AI,
	author = {Aghion, P. and B. F. Jones and C. I. Jones}, 
    year = 2019, 
    title = {{Artificial Intelligence and Economic Growth}}, 
    booktitle = {{The Economics of Artificial Intelligence: An Agenda}},
    editor = {Ajay Agrawal and Joshua Gans and Avi Goldfarb},
    publisher={University of Chicago Press},
    pages={237-282}
}

@article{AcemogluRestrepo2018,
	author = {Acemoglu, Daron and Pascual Restrepo},
    year = 2018, 
    title = {{The Race Between Man and Machine: Implications of Technology for Growth, Factor Shares and Employment}},
    journal = {American Economic Review},
    volume=108,
    issue=6, 
    pages={1488-1542}
}

@book{Harari2017,
	author = {Harari, Yuval Noah},
    year=2017,
    title = {{Homo Deus: A Brief History of Tomorrow}},
    publisher={Vintage}
}

@article{Blasco2023,
  author  = {Blasco, Julien and Guillaud, Elvire and Zemmour, Michael},
  title   = {The Inequality Impact of Consumption Taxes: An International Comparison},
  journal = {Journal of Public Economics},
  year    = {2023},
  volume  = {222},
  pages   = {104897},
  month   = {June},
  doi     = {10.1016/j.jpubeco.2023.104897}
}

@techreport{Berg2021,
  title = {{For the Benefit of All: Fiscal Policies and Equity-Efficiency Trade-offs in the Age of Automation}},
  author = {Berg, Andrew and Bounader, Lahcen and Gueorguiev, Nikolay and Miyamoto, Hiroaki and Moriyama, Kenji and Nakatani, Ryota and Zanna, Luis-Felipe},
  year = {2021},
  month = {July},
  institution = {International Monetary Fund},
  type = {IMF Working Paper},
  number = {WP/21/187},
  url = {https://www.imf.org/en/Publications/WP/Issues/2021/07/02/For-the-Benefit-of-All-Fiscal-Policies-and-Equity-Efficiency-Trade-offs-in-the-Age-of-461152}
}

@article{Harpaz2026,
  title = {Taxing {AI}},
  author = {Harpaz, Assaf},
  journal = {Boston University Law Review},
  volume = {106},
  year = {2026},
  note = {forthcoming},
  url = {https://ssrn.com/abstract=4793836}
}

@techreport{Huynh2025,
  author      = {Huynh, Ky-Cuong and Mittal, Suchet and Frank, Noah},
  title       = {{Funding Government in the Age of AI}},
  institution = {Convergence Analysis},
  year        = {2025},
  month       = {August},
  type        = {Research report}
}

@techreport{Korinek2026,
  title = {{Public Finance in the Age of AI: A Primer}},
  author = {Korinek, Anton and Lockwood, Lee},
  year = {2026},
  month = {February},
  institution = {National Bureau of Economic Research},
  type = {Working Paper},
  number = {34873},
  series = {NBER Working Paper Series},
  address = {Cambridge, MA},
  url = {http://www.nber.org/papers/w34873}
}

@article{Irwin2026,
  title   = {{Token Taxes: Mitigating AGI's Economic Risks}},
  author  = {Irwin, Lucas and Wu, Tung-Yu and Barez, Fazl},
  journal = {arXiv preprint arXiv:2603.04555},
  year    = {2026},
  url     = {https://arxiv.org/abs/2603.04555}
}

@incollection{BrynjolfssonEtal2017,
author = {Brynjolfsson, Erik and Rock, Daniel and Syverson, Chad},
year=2019, 
title={{Artificial Intelligence and the Modern Productivity Paradox: A Clash of Expectations and Statistics}}, 
  booktitle = {{The Economics of Artificial Intelligence: An Agenda}},
    editor = {Ajay Agrawal and Joshua Gans and Avi Goldfarb},
    publisher={University of Chicago Press},
    pages = {23-57}
}

@article{AutorEtal2017,
author = {Autor, David and David Dorn and Lawrence F. Katz and Christina Patterson and Van Reenen, John},
year=2020,
title = {{The Fall of the Labor Share and the Rise of Superstar Firms}}, 
journal = {Quarterly Journal of Economics}, 
volume=135,
issue=2,
pages={645-709}
}

@article{Flug2000,
author = {Flug, Karnit and Hercowitz, Zvi},
year=2000,
title = {{Equipment Investment and the Relative Demand for Skilled Labor: International Evidence}}, 
journal = {Review of Economic Dynamics}, 
volume=3,
issue=3,
pages={461-485}
}

@article{Krusell2000,
author = {Krusell, Per and Ohanian, Lee E. and Ríos-Rull, José-Víctor and Violante, Giovanni L.},
title = {Capital-skill Complementarity and Inequality: A Macroeconomic Analysis},
journal = {Econometrica},
volume = {68},
number = {5},
pages = {1029-1053},
doi = {https://doi.org/10.1111/1468-0262.00150},
url = {https://onlinelibrary.wiley.com/doi/abs/10.1111/1468-0262.00150},
eprint = {https://onlinelibrary.wiley.com/doi/pdf/10.1111/1468-0262.00150},
year = {2000}
}

@book{Tegmark2017,
author = {Tegmark, Max},
year=2017,
title = {{Life 3.0: Being Human in the Age of Artificial Intelligence}},
publisher={New York: Knopf}
}

@book{Bostrom2014,
	title = {{Superintelligence: Paths, Dangers, Strategies}},
	author = {Nick Bostrom},
	publisher = {Oxford University Press},
	year=2014
}

@article{Growiec2020,
	title ={{Automation, Partial and Full}},
	author = {Growiec, Jakub},
	year=2022,
	journal={Macroeconomic Dynamics},
	volume=26,
	issue=7,
	pages = {1731-1755}
}

@article{Growiec2022XS2,
	title ={{What Will Drive Global Economic Growth in the Digital Age?}},
	author = {Growiec, Jakub},
	year=2023,
	journal = {Studies in Nonlinear Dynamics and Econometrics},
	volume=27,
	pages={335-354}
}

@article{AcemogluRestrepo2019b,
	author = {Daron Acemoglu and Pascual Restrepo},
	year=2019,
	title = {{Automation and New Tasks: How Technology Displaces and Reinstates Labor}},
	journal = {Journal of Economic Perspectives},
	volume=33,
	issue=2,
	pages = {3-30}
}

@techreport{HernandezBrown2020,
	author = {Hernandez, D. and Brown, T. B.},
	year=2020,
	title = {{Measuring the Algorithmic Efficiency of Neural Networks}},
	institution={arXiv},
	type = {Preprint arXiv:2005.04305}
}

@incollection{KorinekStiglitz2019,
author = {Korinek, Anton and Stiglitz, Joseph},
   title={{Artificial Intelligence and Its Implications for Income Distribution and Unemployment}}, 
   booktitle = {{The Economics of Artificial Intelligence: An Agenda}},
    editor = {Agrawal, A. and Gans, J. S. and Goldfarb, A.},
   publisher = {University of Chicago Press},
    year = 2019
}

@techreport{GrowiecEtal2024,
	title = {{Hardware and Software: A New Perspective on the Past and Future of Economic Growth}},
	author={Jakub Growiec and Julia Jab{\l}o{\'n}ska and Aleksandra Parteka},
	year=2024, 
	type = {{Working Paper}},
	institution = {Brookings Institution}
}

@misc{epoch2026aimodels,
  title={{AI Models: Our Comprehensive Database of Over 3200 Models Tracks Key Factors Driving Machine Learning Progress}},
  author={{Epoch AI}},
  year=2026,
howpublished = {\url{https://epoch.ai/data/ai-models}},
  note={Accessed: 2026-02-12}
}

@article{owid-artificial-intelligence,
    author = {Charlie Giattino and Edouard Mathieu and Veronika Samborska and Max Roser},
    title = {{Artificial Intelligence}},
    journal = {Our World in Data},
    year = {2023},
    note = {https://ourworldindata.org/artificial-intelligence}
}

@techreport{Karnofsky2016,
	title = {{Some Background on Our Views Regarding Advanced Artificial Intelligence}},
	year=2016,
	author={Holden Karnofsky},
	institution = {Open Philanthropy}
}

@article{Norvig2023,
	title = {{Artificial General Intelligence Is Already Here}},
	author={Ag{\"u}era y Arcas, Blaise and Peter Norvig},
	year=2023,
	journal = {Noema},
	pages = {October 10, 2023}
}

@book{YudkowskySoares2025,
	author={Eliezer Yudkowsky and Nate Soares},
	year=2025,
	title = {{If Anyone Builds It, Everyone Dies}},
	publisher = {Little, Brown and Company}
}

@article{GrowiecPrettner2025,
	author = {Jakub Growiec and Klaus Prettner},
	title = {{The Economics of p(doom): Scenarios of Existential Risk and Economic Growth in the Age of Transformative AI}},
	year=2026,
	journal = {Economic Modelling},
    pages={107718},
    volume=163
}

@article{NatureAGI,
author={Keming Chen, Eddy and Mikhail Belkin and Leon Bergen and David Danks},
title={{Does AI Already Have Human-Level Intelligence? The Evidence Is Clear}},
year=2026,
journal={Nature},
volume=650,
pages={36-40}
}

@book{Aguera2025,
author={Ag{\"u}era y Arcas, Blaise},
title = {{What Is Intelligence?}},
year=2025,
publisher={MIT Press, Cambridge MA}
}

@techreport{TrammellPatel2025,
title={{Capital in the 22nd Century}},
author={Philip Trammell and Dwarkesh Patel},
year=2025,
institution={Stanford Digital Economy Lab},
type={https://philiptrammell.substack.com/p/capital-in-the-22nd-century}
}

@misc{METR,
  title = {{Measuring AI Ability to Complete Long Tasks}},
  author = {METR},
  howpublished = {\url{https://metr.org/blog/2025-03-19-measuring-ai-ability-to-complete-long-tasks/}},
  year = {2025},
  month = {03},
  }

@techreport{CritchRussell2023,
	title = {{TASRA: a Taxonomy and Analysis of Societal-Scale Risks from AI}},
	year=2023,
	author={Andrew Critch and Stuart Russell},
	institution={arXiv},
	type ={2306.06924}
}

@article{MinnitiEtal2025,
title = {{AI Innovation and the Labor Share in European Regions}},
journal = {European Economic Review},
volume = {177},
pages = {105043},
year = {2025},
issn = {0014-2921},
doi = {https://doi.org/10.1016/j.euroecorev.2025.105043},
url = {https://www.sciencedirect.com/science/article/pii/S0014292125000935},
author = {Antonio Minniti and Klaus Prettner and Francesco Venturini}
}

@book{AgrawalEtal2022,
    title = {{Power and Prediction: The Disruptive Economics of Artificial Intelligence}},
    year=2022,
    author = {Ajay Agrawal and Joshua Gans and Avi Goldfarb},
    Publisher ={Harvard Business Review Press}
}

@techreport{KulveitEtal2025,
    title = {{Gradual Disempowerment: Systemic Existential Risks from Incremental AI Development}},
    author = {Jan Kulveit and Raymond Douglas and Nora Ammann and Deger Turan and David Krueger and David Duvenaud},
    type={{arXiv:2501.16946}},
    institution={ACS Research Group},
    year=2025
}

@article{Dung2025,
	title = {{The Argument for Near-Term Human Disempowerment Through AI}},
	author = {Leonard Dung},
	year=2025,
	journal={AI and Society},
  	pages = {1195-1208},
    volume=40
}

@techreport{AI2027,
    title = {{AI 2027}},
    author = {Daniel Kokotajlo and Scott Alexander and Thomas Larsen and Eli Lifland and Romeo Dean},
    year=2025,
    institution={AI Futures Project},
    type = {ai-2027.com}
}

@techreport{Amodei2026,
    title = {{The Adolescence of Technology}},
    author = {Dario Amodei},
    year=2026,
    institution={Anthropic},
    url={https://www.darioamodei.com/essay/the-adolescence-of-technology},
    type = {Essay}
}

@techreport{AgrawalEtal2025,
title = {{The Economics of Bicycles for the Mind}},
author = {Ajay K. Agrawal and Joshua S. Gans and Avi Goldfarb},
type = {{Working Paper 34034}},
year=2025,
institution = {{NBER}}
}

@techreport{KorinekSuh2024,
    title={{Scenarios for the Transition to AGI}},
    author={Anton Korinek and Donghyun Suh},
    year=2024,
    institution={National Bureau of Economic Research},
    type = {{Working Paper 32255}}
}

@techreport{Jablonska2025,
	author={Jab{\l}o{\'n}ska, Julia},
	year=2025,
	title = {{On the Substitutability Between Automation Technologies and Human Cognitive Work}},
	type = {Ph.D. Dissertation},
	institution = {{SGH Warsaw School of Economics}}
}

@article{GrowiecEtal2026,
title = {{Workers’ Incentives and the Optimal Taxation of AI}},
journal = {Economics Letters},
volume = {266},
pages = {113062},
year = {2026},
issn = {0165-1765},
doi = {https://doi.org/10.1016/j.econlet.2026.113062},
author = {Jakub Growiec and Klaus Prettner and Maciej Szkróbka}
}

@book{Bastani2019,
    title = {{Fully Automated Luxury Communism: A Manifesto}},
    author = {Aaron Bastani},
    publisher = {Verso Books},
    year=2019
}

@book{Roemer1994,
 title = {{A Future for Socialism}},
 author={John E. Roemer},
 year=1994,
 publisher={Harvard University Press}
}

@book{AI2040,
    title = {{AI 2040: Plan A}},
    author = {Thomas Larsen and Romeo Dean and Brendan Halstead and Eli Lifland and Ryan Greenblatt and Daniel Kokotajlo},
    year=2026,
    publisher = {AI Futures Project}
}

@article{PikettyEtal2023,
 title = {{Rethinking Capital and Wealth Taxation}},
 author = {Thomas Piketty and Emmanuel Saez and Gabriel Zucman},
 year=2023,
 journal = {Oxford Review of Economic Policy}, volume=39,
 pages={575-591}
}

@techreport{NormalTech,
   title = {{AI as Normal Technology: 
An Alternative to the Vision of AI as a Potential Superintelligence}},
    author = {Arvind Narayanan and Sayash Kapoor},
    year=2025,
    type = {Essay},
    institution = {Knight First Amendment Institute}
}

@article{BonfiglioliEtal2025, 
	author={Bonfiglioli, Alessandra and Crin\`o, Rosario and Gancia, Gino and Papadakis, Ioannis},
	year=2025,
	title={{Artificial Intelligence and Jobs: Evidence from US Commuting Zones}},
	journal={Economic Policy},
	volume=40,
	issue=121,
	pages={145-194}
}

@article{Moll2022,
  title = {Uneven {{Growth}}: {{Automation}}'s {{Impact}} on {{Income}} and {{Wealth Inequality}}},
  shorttitle = {Uneven {{Growth}}},
  author = {Moll, Benjamin and Rachel, Lukasz and Restrepo, Pascual},
  year = 2022,
  journal = {Econometrica},
  volume = {90},
  number = {6},
  pages = {2645--2683},
  issn = {0012-9682},
  doi = {10.3982/ECTA19417},
  urldate = {2025-12-22},
  langid = {english}
}

@book{PrettnerBloom2020_adapted,
  title = {Automation and {{Its Macroeconomic Consequences}}: {{Theory}}, {{Evidence}}, and {{Social Impacts}}.},
  author = {Prettner, K. and Bloom, D.},
  year = 2020,
  publisher = {Academic Press, Amsterdam, NL.}
}

@article{Berg2018,
  title = {Should We Fear the Robot Revolution? ({{The}} Correct Answer Is Yes)},
  author = {Berg, A and Buffie, A and Zanna, L.-F.},
  year = 2018,
  journal = {Journal of Monetary Economics},
  volume = {Vol. 97},
  number = {No. C},
  pages = {117--148}
}

@article{Corneo2018,
  title = {Ein {{Staatsfonds}}, Der Eine Soziale {{Dividende}} Finanziert},
  author = {Corneo, G},
  year = 2018,
  journal = {Perspektiven der Wirtschaftspolitik},
  volume = {Vol. 19},
  number = {No. 2},
  pages = {94--109}
}

@techreport{BernhardFiedler2025,
  title = {Basic {{Income}} and {{Labor Supply}}: {{Evidence}} from an {{RCT}} in {{Germany}}},
  type={{{Discussion Papers}} of {{DIW Berlin}} 2123},
  institution={{{DIW Berlin}}, {{German Institute}} for {{Economic Research}}.},
  author = {Bernhard, Sarah and Bohmann, Sandra and Fiedler, Susann and Kasy, Maximilian and Schupp, J{\"u}rgen and Schwerter, Frederik},
  year = 2025
}

@article{gersbachArtificialIntelligenceSelflearning2025,
  title = {Artificial Intelligence as Self-Learning Capital},
  author = {Gersbach, Hans and Komarov, Evgenij and Von Maydell, Richard},
  year = 2025,
  month = dec,
  journal = {Economic Modelling},
  volume = {153},
  pages = {107221},
  issn = {02649993},
  doi = {10.1016/j.econmod.2025.107221},
  urldate = {2026-06-12},
  langid = {english}
}

@article{gilbertTariffsTaxesWages2026,
  title = {Tariffs, Taxes and Wages in the Age of Artificial Intelligence},
  author = {Gilbert, John and Koska, Onur A. and Oladi, Reza},
  year = 2026,
  month = apr,
  journal = {Economic Inquiry},
  volume = {64},
  number = {2},
  pages = {395--408},
  issn = {0095-2583, 1465-7295},
  doi = {10.1111/ecin.70048},
  urldate = {2026-06-12},
  langid = {english}
}

@article{Gasteiger2026,
  title = {Electricity {{Use}} of {{Automation}} or {{How}} to {{Tax Robots}}?},
  author = {Gasteiger, Emanuel and Kuhn, Michael and Mistlbacher, Matthias and Prettner, Klaus},
  year = 2026,
  month = feb,
  journal = {Scottish Journal of Political Economy},
  volume = {73},
  number = {1},
  pages = {e70032},
  issn = {0036-9292, 1467-9485},
  doi = {10.1111/sjpe.70032},
  urldate = {2026-02-03},
  langid = {english}
}

@misc{Economist2018a,
  title = {On Firmer Ground: {{The}} Time May Be Right for Land-Value Taxes. {{August}} 9th, 2018},
  author = {{The Economist}},
  year = 2018
}

@article{Frank2008,
  title = {Progressive {{Consumption Tax}}},
  author = {Frank, R},
  year = 2008,
  journal = {Democracy: A Journal of Ideas},
  volume = {Vol. 8}
}

@article{Acemoglu2025Simple,
  title = {The Simple Macroeconomics of {{AI}}},
  author = {Acemoglu, Daron},
  year = 2025,
  month = jan,
  journal = {Economic Policy},
  volume = {40},
  number = {121},
  pages = {13--58},
  issn = {0266-4658, 1468-0327},
  doi = {10.1093/epolic/eiae042},
  urldate = {2025-05-20},
  copyright = {https://academic.oup.com/pages/standard-publication-reuse-rights},
  langid = {english}
}

@techreport{KorinekEtal2026,
 title = {{Economic Scenarios for Transformative AI}},
 author = {Anton Korinek and Charles I. Jones and  Szymon Sacher and Tess Cotter and Peter McCrory},
 institution={The Anthropic Institute},
 type={{Working Paper No. 2026-02}},
 year=2026
}

\end{document}